\documentclass{aa}
\usepackage{epsfig,times}
\usepackage{amssymb}
\usepackage{graphicx}
\begin{document}

\title{Anisotropic convection in rotating proto-neutron stars}
\author{J.~A.~Miralles \inst{1} \and J.~A.~Pons \inst{1} \and V.~ Urpin \inst{1,2}}
\institute{Departament de F\'{\i}sica Aplicada, Universitat d'Alacant,
           Ap. Correus 99, 03080 Alacant, Spain 
\and A.F.Ioffe Institute of Physics and Technology and Isaak Newton Institute
of Chili, Branch in St.Petersburg, 194021 St.Petersburg, Russia}

\date{Received...../ Accepted.....}

\abstract{We study the conditions for convective instability in rotating,
non-magnetic proto--neutron stars. The criteria that determine stability
of nascent neutron stars are analogous to the Solberg--H{\o}iland conditions 
but including the presence of lepton gradients. Our results show that, 
for standard angular velocity profiles, convectively unstable modes with 
wave-vectors parallel to the rotation axis are suppressed by a stable 
angular momentum profile, while unstable modes with 
wave-vectors perpendicular to the axis remain unaltered. Since the wave-vector
is perpendicular to the velocity perturbation, the directional
selection of the unstable modes may result in fluid motions along 
the direction of the rotation axis. This occurs in  rigidly rotating stars 
as well as in the inner core of differentially rotating stars.  
Our results provide a natural source of asymmetry for proto--neutron stars 
with the only requirement that angular velocities be
of the order of the convective characteristic frequency. 

}

\titlerunning{Anisotropic convection in rotating PNSs}
\authorrunning{Miralles, Pons, \& Urpin}

\maketitle

\section{Introduction}

Hydrodynamical instabilities in newly born neutron stars play an important 
role in enhancing neutrino luminosities and increasing the energy deposition 
efficiency. It was first argued by Epstein (1979) that the negative lepton 
gradient that arises in the outer layers of the proto-neutron star (PNS) can 
be convectively unstable. However, convection in PNSs can be driven not only 
by the lepton gradient but by a negative entropy gradient as well (see, e.g., 
Arnett 1987, Burrows \& Lattimer 1988). The development of negative entropy 
and lepton gradients is common in many simulations of supernova models 
(Bruenn \& Mezzacappa 1994, Bruenn et al. 1995) and evolutionary models of 
PNSs (Burrows \& Lattimer 1986, Keil \& Janka 1995, Sumiyoshi et al. 1995, 
Pons et al. 1999). 
The nature of instabilities in PNSs has been considered by a number of 
authors (Grossman et al. 1993; Bruenn \& Dineva 1996; 
Miralles et al.  \cite{MPU00}, MPU in the following; 
Miralles et al.  \cite{MPU02}). 
In non-rotating and non-magnetic PNSs, the criteria indicate the 
presence of two essentially different instabilities (MPU)
with a convectively unstable region 
surrounded by a neutron-finger unstable region, the latter involving 
typically a larger portion of the stellar material. The unstable zones 
grow in size and move inward as the PNS deleptonizes (Keil et al. 1996), 
until neutrino diffusion progressively reduces the temperature and lepton 
gradients, and convective instabilities disappear, at most in 30-40 s. The 
growth time of convective and neutron finger instabilities are substantially 
different, as well as the efficiency of turbulent transport in the convective 
and neutron-finger unstable zones.

Less is known about convection in rotating PNSs. The failure of detailed 
simulations in spherical symmetry to get a successful Supernova explosion 
has led different authors to argue that multidimensional effects, such 
as convection or  rapid rotation can be the key to the problem (see e.g. 
Buras et al. 2003, and references therein). From theoretical modeling and 
simple analytic considerations it is commonly accepted that core collapse 
of a rotating progenitor leads to differential rotation of a newly born 
neutron star (Zwerger \& M\"uller 1997, Rampp et al. 1998, Liu 2002, 
Dimmelmeier et al. 2002, M\"uller et al. 2003), mainly due to conservation 
of the angular momentum during collapse. A recent study on evolutionary 
sequences of rotating PNSs (Villain et al. \cite{Loic03}) shows that the 
typical scale on which the angular velocity changes is in the range 
$\approx 5-10$ km. 
Granted that a PNSs is born rapidly rotating, when the angular 
velocity is of the order or larger than the Brunt-V\"ais\"al\"a or Ledoux 
frequencies, even relatively fast convective instabilities can be 
substantially modified, and convection can be constrained to the polar 
regions (Fryer \& Heger \cite{FH00}). In this paper, we study this effect 
in the context of PNSs.

\section{The dispersion relation.}
Consider a PNS rotating with  angular velocity $\Omega = \Omega(s, z)$. 
In what follows we use cylindrical coordinates ($s$, $\varphi$, $z$). 
The deleptonization and cooling timescales are assumed to be much longer 
than the growth time of instability, thus it can be treated in a 
quasi-stationary approximation. For the sake of simplicity, we consider 
axisymmetric short-wavelength perturbations with spatial and temporal dependence 
$\exp(\gamma t - i \vec{k} \cdot \vec{r})$ where $\vec{k}= (k_{s}, 0, k_{z})$ 
is the wave-vector. Small perturbations will be indicated by a subscript 1, 
whilst unperturbed quantities will have no subscript, except when indicating 
vector components. In the unperturbed state, the PNS is in hydrostatic 
equilibrium,
\begin{equation}
\frac{\nabla p}{\rho} = \vec{G} \;, \;\;\;
\vec{G} = \vec{g} + \Omega^{2} \vec{s}
\label {eqh}
\end{equation}
where $\vec{g}$ is the gravity. 

In the Boussinesq approximation, the linearized momentum and continuity 
equations read
\begin{equation}
\gamma \vec{v}_{1} + 2 \vec{\Omega} \times \vec{v}_{1} +
\vec{e}_{\varphi} s (\vec{v}_{1} \cdot \nabla \Omega) =
\frac{i \vec{k} p_{1}}{\rho} + \vec{G} \; \frac{\rho_{1}}{\rho} \;,
\end{equation}
\begin{equation}
\vec{k} \cdot \vec{v}_{1} = 0 \;, 
\end{equation}
where $\vec{v}_{1}$, $p_{1}$ and $\rho_{1}$ are perturbations of the 
velocity, pressure and density, respectively; $\vec{e}_{\varphi}$ is 
the unit vector in the azimuthal direction. We assume that the matter is 
in chemical equilibrium, thus the density is generally a function of the 
pressure $p$, temperature $T$ and lepton fraction $Y$. In the Boussinesq 
approximation, perturbations of the pressure are small, therefore $\rho_1$ 
can be expressed in terms of the perturbations of temperature, $T_1$, 
and lepton fraction, $Y_1$,
\begin{equation}
\rho_{1} \approx - \rho \left( \beta \frac{T_{1}}{T} + \delta Y_{1} 
\right),
\end{equation} 
where $\beta$ and $\delta$ are the coefficients of thermal and chemical
expansion; $\beta = - (\partial \ln \rho/\partial \ln T)_{pY}$, $\delta =
- (\partial \ln \rho/ \partial Y)_{pT}$.

We consider instabilities arising on a timescale much shorter than the 
dissipative timescale, i.e., those perturbations
with relatively long wavelengths but still smaller than the pressure
scale, to ensure that our approximation is valid.
For example, perturbations with $\lambda =
2 \pi /k \sim 0.3-1$ km satisfy both these requirements. For such 
perturbations, we can neglect dissipative effects in the equations 
governing thermal balance and lepton fraction (see MPU for more details 
about the influence of dissipative terms). Then, the linearized transport 
equations read
\begin{equation}
{\dot{T}_{1}} - \vec{v}_{1} \cdot  {\Delta\nabla T} = 0,
\label{temp} 
\end{equation}
\begin{equation}
\dot{Y}_{1} + \vec{v}_{1} \cdot \nabla Y = 0,
\label{lept} 
\end{equation}
where $\Delta \nabla T$ is the super-adiabatic temperature gradient:
$$
\Delta\nabla T
=  \left( \frac{\partial T}{\partial p} \right)_{s,Y} \nabla p
-  \nabla T~.
$$

The dispersion equation corresponding to Eqs. (\ref{eqh})--(\ref{lept}) is
\begin{equation}
\gamma^{2} (\gamma^{2} + q^{2} - \omega_{g}^{2} - \omega_{L}^{2}) = 0,
\label{disp0}
\end{equation}
where 
\begin{eqnarray}
&& q^{2} = \frac{k_{z}^{2}}{k^{2}} \; \Omega_{e}^{2} - \frac{k_{s} k_{z}}{k^{2}}
{\Omega^{2}_z},
\nonumber \\  
&& \omega_{g}^{2} = - (\beta/T) \vec{G}_\perp \cdot \Delta\nabla T~ ,
\quad \quad
\omega_{L}^{2} = \delta \vec{G}_\perp \cdot \nabla Y ,        
\nonumber
\end{eqnarray}
and $\vec{G}_\perp = \vec{G} - \vec{k} ( \vec{k} \cdot \vec{G})/k^{2}$; $\Omega_{e}$ 
is the epicyclic frequency, $\Omega_{e}^{2} = \frac{1}{s^3} \partial (s^{4} 
\Omega^{2})/ \partial s$, and we have defined $\Omega_{z}^{2} =  s~\partial 
\Omega^{2}/\partial z$. Eq. (\ref{disp0}) differs from the standard dispersion 
equation of buoyant waves in a differentially rotating fluid (see, e.g., Goldreich 
\& Schubert 1967) by the presence of a term proportional to $\nabla Y$.

Non-trivial solutions of Eq. (\ref{disp0}) are given by
\begin{equation}
\gamma^{2} = - q^{2} + \omega_{g}^{2} + \omega_{L}^{2}.
\end{equation} 
The condition of instability is $~ \omega_{g}^{2} + \omega_{L}^{2} 
> q^{2}$. In non--rotating PNSs, the characteristic frequencies 
in the convective regions are typically of the order of 1 ms$^{-1}$ (MPU).
Therefore, rotational effects will be important when the angular
velocity $\Omega$ is of the same order, $\Omega \approx 1000$ rad/s,
which can be reached during the early stages of PNS evolution
(Villain et al., \cite{Loic03}).

\section{The instability criteria.} 

Assume that a vector $\vec{k}$ forms 
an angle $\phi$ with the polar axis. Defining the following vector,
\begin{eqnarray}
\vec{C} = - (\beta/T) \Delta\nabla T + \delta \nabla Y~ ,
\end{eqnarray}
the dispersion relation can be written in terms of the
components of $\vec{C}$ and $\vec{G}$ as follows:
\begin{equation}
\gamma^2 = A~{\cos^2{\phi}} + B~{\sin{\phi} \cos{\phi}}
+ C_z G_z
\label{disp1}
\end{equation}
where $ A = -\Omega_{e}^{2} + C_s G_s - C_z G_z$ and $B = \Omega_{z}^{2} 
- C_z G_s -C_s G_z$. By taking the curl of Eq. (\ref{eqh}),
it can be readily obtained that the condition of hydrostatic equilibrium
leads to 
\begin{equation}
\Omega_{z}^{2} = \left[\vec{C}\times\vec{G}\right]_{\varphi} =C_z G_s -C_s G_z
\label{eqh2}
\end{equation}
so that $B$ can be further simplified to obtain $B = - 2 C_s G_z$. 

Since the dependence of Eq. (\ref{disp1})  on $\phi$ is rather simple, we 
can obtain that the maximum of $\gamma^2$ corresponds to
\begin{equation}
\cos^{2}\phi = \frac{1}{2} 
\left[ 1 \pm \sqrt{\frac{A^{2}}{A^2 + B^2}} \right]~,
\label{phimax}
\end{equation}
with the $\pm$ sign depending on the sign of $A$,
and the maximum value of $\gamma^2$ is given by
\begin{eqnarray}
\gamma_{max}^{2}  &=&\frac{-\Omega_{e}^{2} - \omega_0^2}{2} +
\nonumber \\
&& + \frac{1}{2}\sqrt{\left[ \Omega_{e}^{2} + \omega_0^2 \right]^2
+ 4 G_z (C_z \Omega_{e}^{2} - C_s \Omega_{z}^{2}) }.
\nonumber
\end{eqnarray}
where $\omega_0^2 = - (\vec{C}\cdot\vec{G})$. The two conditions for 
stability follow straightforwardly from the above expression:
\begin{eqnarray}
-\Omega_{e}^{2} - \omega_0^2 < 0~; \quad \quad
G_z (C_z \Omega_{e}^{2} - C_s \Omega_{z}^{2}) < 0~,
\label{cond2} 
\end{eqnarray}
These two conditions look like the Solberg--H{\o}iland conditions
(Tassoul \cite{Tas00}), but with additional terms due to the lepton 
gradients. Criteria (\ref{cond2}) can be also applied to
quasi-Keplerian toroidal configurations made of nuclear matter
surrounding black holes, in which the effect of lepton gradients
is relevant.

Notice that if isobaric and isopicnic surfaces coincide, $\vec{C}$ and
$\vec{G}$ are aligned, and vertical gradients of $\Omega$ are not 
allowed. Consistently, most of the studies of rotating stars with
simplified polytropic equations of state use angular rotation
profiles that depend only on $s$. In our calculations, we use the
following law proposed by Komatsu et al. (\cite{keh89}) and used by
many authors:
\begin{equation}
\Omega(s) = \frac{\Omega_c R_0^2}{s^2 + R_0^2}.
\label{slaw}
\end{equation}

\begin{figure}
\resizebox{\hsize}{!}{\includegraphics{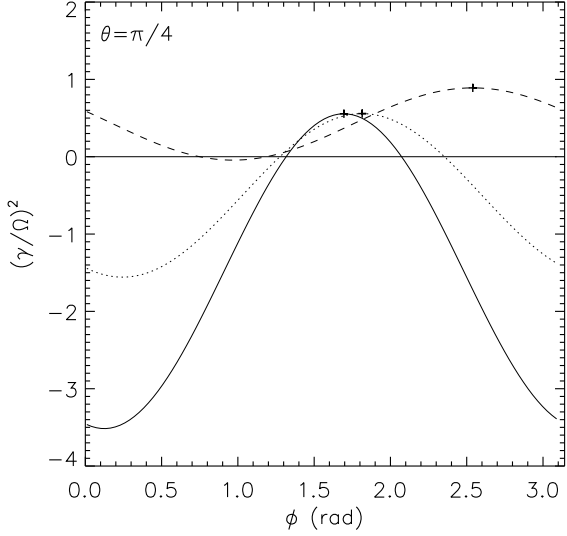}}
\caption{Characteristic squared frequency normalized to
the rotation frequency, $\gamma^2/\Omega^2$, as a function of
the angle between the wavevector and the polar axis ($\phi$),
assuming that $\omega_0^2=\Omega^2$. The rotation law is 
described by Eq. (\ref{slaw}). Three different radial
points, with $s/R_0=0.1, 1, $and 5 are plotted with solid lines,
dots, and dashes, respectively.
}
\label{fig1}
\end{figure}

In Fig. 1 we show $\gamma^2/\Omega^2$ as a function of the wavevector
orientation in a convectively unstable situation, $\omega_0^2>0$. We have taken
$\omega_0^2=\Omega^2$, and the polar angle $\theta=\pi/4$. For simplicity, 
we have used the law (\ref{slaw}) for the angular velocity so that 
$\Omega_z=0$. Three different cases are plotted with solid lines, dots, 
and dashes, corresponding to $s/R_0=0.1, 2, $ and 5, respectively.
When $s/R_0 \ll 1$ the angular velocity varies slowly and the situation 
is close to rigid rotation, while for $s/R_0 \gg 1$ the specific angular 
momentum is nearly constant and rotational effects are negligible.
When the angular momentum increases with the distance to the axis
(Rayleigh stable), $\gamma^2$ can become negative and instability is 
suppressed for a subset of the possible orientations of $\vec{k}$. Only 
perturbations with ${\vec k}$ oriented more or less perpendicular to the 
polar axis (fluid motions in the vertical direction) are not suppressed. 
Note that the fact that rotation results in turbulent convective motions 
preferentially in the direction parallel to the rotation axis was first
pointed out by Randers (\cite{Ran42}) in the context of the study of 
energy transport in stellar convective cores.

\begin{figure}
\resizebox{\hsize}{!}{\includegraphics{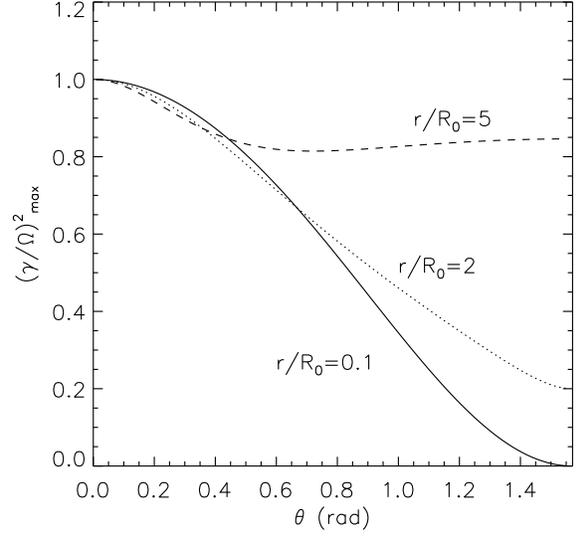}}
\caption{Maximum of $\gamma^2/\Omega^2$ as a function of
the polar angle for a fixed radius of $r/R_0=0.1$ (solid), 
2.0 (dots) and 5.0 (dashes).
}
\label{fig2}
\end{figure}

\begin{figure}
\resizebox{\hsize}{!}{\includegraphics{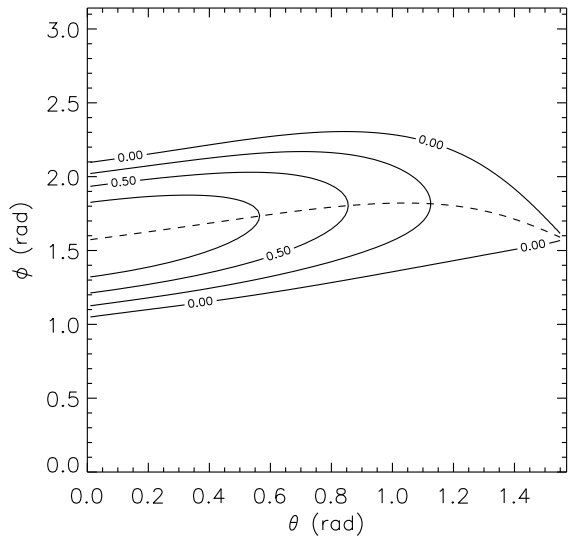}}
\caption{Surfaces of constant $\gamma^2/\Omega^2$ in a
$\theta-\phi$ plane, for  $r/R_0=1$. The region enclosed by the 
surface  $\gamma^2/\Omega^2=0$ is convectively unstable. The dashed
line indicates the angle at which the maximum is reached.
}
\label{fig3}
\end{figure}

This effect is more manifest looking at Figs. \ref{fig2} and \ref{fig3}. 
In Fig. 2 we plot the maximum value of $\gamma^2$ as a function of the 
polar angle ($\theta$), at three different spherical radii. Near the pole
the maximum of $\gamma$ does not change, but as we approach the equator, 
rotational effects may reduce its value drastically, unless the gradient 
of the angular momentum is small (case $r/R_0=5$). More importantly, not 
only does the growth time become longer, but the unstable perturbations are 
restricted to some range of angles, centered at the value given by 
Eq. (\ref{phimax}), and width given by
\begin{equation}
\tan{\Delta \phi}= 
\frac{\sqrt{- G_z (C_z \Omega_{e}^{2} - C_s \Omega_{z}^{2})}}
{(-\Omega_{e}^{2} - \omega_0^2)}
\end{equation}
In Fig. \ref{fig3} we plot surfaces of constant $\gamma^2/\Omega^2$ 
in a $\theta-\phi$ plane, for  $r/R_0=1$. The curve $\gamma^2=0$
encloses the interval of angles in which perturbations are unstable.
The position of the maximum is indicated by the dashed line.
Notice that when the second of the stability conditions (\ref{cond2}) is 
not fulfilled, the function $\gamma^2$ has no zeros. Therefore, $\gamma^2$ is
either positive or negative for all angles depending on the sign of 
the first condition.

\section{Discussion. }
\begin{figure}
\resizebox{\hsize}{!}{\includegraphics{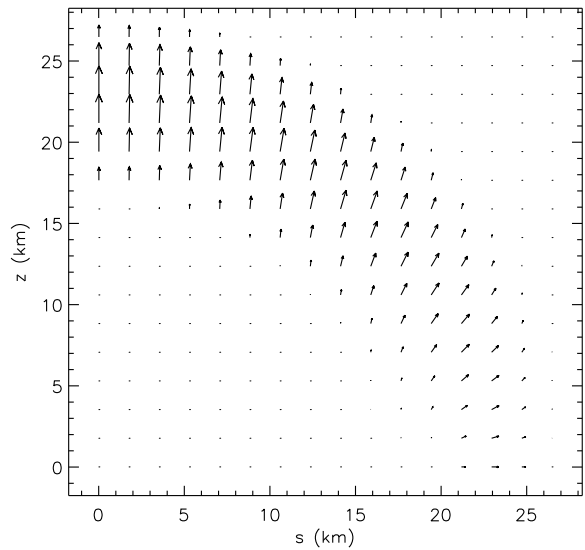}}
\caption{The direction of the motion for the most rapidly
growing modes in a realistic model of PNS 0.5 s after birth. In the
convective region (17-28 km), the instability is more effective in the polar
region and, even at intermediate latitudes, vertical motions are preferred.
}
\label{fig4}
\end{figure}

We have derived the criteria for convective instability in rotating,
non--magnetic PNSs, obtaining two conditions analogous to the 
Solberg--H{\o}iland criteria but including the effects of lepton 
gradients. In convectively unstable regions, the existence of a positive
gradient of angular momentum can stabilize the unstable convective modes
in some directions. The basic picture is that an unstable buoyant element 
of fluid is stabilized when it tries to move toward a direction in which the 
surrounding fluid elements have larger angular momentum, because it is subject to
a smaller centrifugal force and therefore restored to its equilibrium position.
However, this restoring force has no effect on buoyant waves that follow
lines of constant angular momentum. For example, in a situation in which 
thermal, lepton, and pressure gradients are nearly aligned, the angular momentum 
depends weakly on the vertical coordinate, thus allowing convectively 
unstable modes to grow only in the direction approximately parallel to the 
rotation axis. In a general case, the geometry will be more complicated but 
the qualitative results will not change. In Fig. \ref{fig4} we illustrate 
this effect in a realistic model of a PNS 0.5 s after birth, taken from 
Pons et al. (1999). The convectively unstable region extends from 17 to 28 km, 
and the arrows indicate the direction of the most rapidly growing mode, with 
the length being proportional to the inverse growth time. 
In the convectively unstable region near the equator, $\Omega_e$ varies 
between 500 and 1000 rad/s, which is about one third the value of $\omega_0$,
and we have taken $R_0=10$ km (defined in Eq. (\ref{slaw})).
In a saturated state, hydrodynamical non-linear effects will 
probably smear out anisotropies on small scales but it is unlikely that this 
may occur on the largest convective scale.  Hydrodynamic simulations of the 
early evolution of rotating PNSs show that the scenario drawn in Fig. \ref{fig4} 
is fulfilled during the very early phase of convection in PNSs 
(H-Th. Janka, private communication), which shows that our linear
stability analysis reflects accurately the situation at the onset of convection.
However, after several hundreds of milliseconds, the efficient neutrino and 
convective transport in the polar region will flatten the temperature, lepton
and angular momentum profiles, resulting in a different situation: convection
occurring only close to the equator, although convective shells are still elongated
and parallel to the rotation axis (see e.g. Fig. 10 in Janka \& Keil \cite{JK97},
or Janka, Kifonidis \& Rampp \cite{JKR01} ).

This coupling between convective instability and stabilizing gradients of 
angular momentum can be another ingredient that helps to explain the large asymmetry that 
has been observed in recent simulations of rotational core collapse Supernovae 
(M\"uller et al., 2003), not observed in simulations of convective PNSs without 
rotation.  The mechanism we propose is a natural way to create anisotropic 
energy and momentum transport by convective motions, that only requires
that the angular velocity be of the order of the Brunt--V\"ais\"al\"a or leptonic 
frequencies. Notice that our results are relevant for 
convective motions in the PNS, more than in the extended envelope as in
Fryer \& Heger (\cite{FH00}). Our analysis is only valid in a steady--state
situation, while the conditions in the accretion layer are extremely non--stationary,
and other effects such as anisotropic accretion are certainly more relevant.
If the collapsed core shows strong differential 
rotation, at scales larger than 50-100 km the angular momentum can be nearly constant
and rotational effects are small. 
Nevertheless, the anisotropy originated in the inner core may have some effect
at larger scales, and can help to explain the Supernova mechanism.
Obviously, self--consistent 3D simulations involving convection,
neutrino transport and angular momentum transport are needed prior to drawing
robust conclusions.

In our study we only considered non--magnetic proto--neutron stars. The presence 
of a large magnetic field could change this picture.  It has recently been shown 
that turbulent mean--field dynamo can be effective for PNSs with periods 
shorter than 1 s, which would generate strong magnetic fields in the 
interior (Bonanno et al. \cite{BRU}). However, magnetic field effects begin 
to be relevant only when the Alfven frequency is of the order of the other 
dynamical or dissipative frequencies, and this requires magnetic fields larger 
than 10$^{15}$ G (Miralles et al. \cite{MPU02}). Probably the missing ingredient 
to explain Supernovae is related to the coupling between different effects, such 
as the one proposed in this paper, the magneto--rotational instability (Akiyama et al., 
\cite{Aki03}), or a combination of all of them.

\section*{Acknowledgments}
This work has been supported by the Spanish Ministerio
de Ciencia y Tecnolog\'{\i}a grant AYA 2001-3490-C02.
JAP is supported by a {\it Ram\'on y Cajal} contract from the Spanish MCyT.
VU is supported by the Russian Foundation of Basic Research (04-02-16243) and
a grant from Generalitat Valenciana. We thank the referee of this paper
(H.-Th. Janka) for useful comments and sharing unpublished results with us.


\end{document}